\documentclass[12pt]{article}

\newcommand{\beq}{\begin{eqnarray}}
\newcommand{\eeq}{\end{eqnarray}}
\newcommand{\rf}[1]{(\ref{#1})}

\begin{document}


\begin{center}
\hspace{10cm}

\begin{flushright}
BCCUNY-HEP/02-03
\end{flushright}

\vspace{15pt}
{\large \bf
GRIFFITHS INEQUALITIES FOR SOME O($n$) CLASSICAL SPIN MODELS
WITH $n\ge 3$ }

\vspace{15pt}
\vspace{15pt}

\vspace{20pt}

{\bf Peter Orland}$^{\rm a.}$$^{\rm b.}$\footnote{Work supported in
part by National Science Foundation Grant 0070991 and PSC-CUNY Research Award 63502-00 32.}

\vspace{8pt}
 
\begin{flushleft}
a. Physics Program, The Graduate School and University Center,
The City University of New York, 365 Fifth Avenue,
New York, NY 10016, U.S.A.
\end{flushleft}

\begin{flushleft}
b. Department of Natural Sciences, Baruch College, The 
City University of New York, 17 Lexington Avenue, New 
York, NY 10010, U.S.A., orland@gursey.baruch.cuny.edu 
\end{flushleft}

\vspace{40pt}

{\bf Abstract}
\end{center}

\noindent
The first and second Griffiths inequalities 
are proved for 
some classical O($n$)-invariant spin models (including Euclidean quantum field theories)
for any $n$. The proof assumes a certain condition on an integral transform
of the measure. Some examples are discussed.

\vspace{30pt}

\section{Introduction}
\setcounter{equation}{0}
\renewcommand{\theequation}{1.\arabic{equation}}

We consider in this paper classical spin models on a lattice $\Lambda$, with an
$n$-component vector
spin $s(x)=(s_{1}(x),\dots s_{n}(x))$, assigned to
each lattice site $x\in \Lambda$. Such a model has the partition function
\beq
Z[G]=\int \left\{\prod_{x\in \Lambda} d^{n}s(x)\; \mu[s(x)] \right\}
\exp \left[\sum_{x,y \in \Lambda} J_{xy}\;s(x)\cdot s(y)+\sum_{x\in \Lambda} 
h_{x} \cdot s(x)\right]\;, \label{partition}
\eeq
where $\mu(s)=\mu(s_{1}, \dots,s_{n} )\ge 0$ is 
a nonnegative distribution on ${{\rm I} \! {\rm R}}^{n}$ 
and
$J_{xy}=J_{yx}$ and $h_{x}$ are real nonnegative numbers for 
each pair $x,y\in \Lambda$. The
expectation values are
\beq
\left< s_{q_{1}}(x_{1})\cdots s_{q_{m}}(x_{m})\right>
=\frac{1}{Z[G]}
\int \left\{\prod_{x\in \Lambda} d^{n}s(x)\; 
\mu[s(x)] \right\}
\nonumber
\eeq
\beq
\times\;s_{q_{1}}(x_{1})\cdots s_{q_{m}}(x_{m})\;
\exp \sum_{x,y \in \Lambda} J_{xy}\;s(x)\cdot s(y)
\;, \label{correl}
\eeq
Some
special cases possess
global O($n$) symmetry, but in general such a symmetry may not be 
present in \rf{partition}
and \rf{correl}. If the measure $\mu(s)$ depends only on 
$s\cdot s$ an O($n$) symmetry is realized through 
$s(x)\rightarrow Ms(x)$, where $M$ is an
arbitrary 
$n\times n$ orthogonal matrix, $M^{T}=M^{-1}$ whose components 
are real. 

The G.K.S. inequalities were first proved by Griffiths 
\cite{griffiths1,griffiths2,griffiths3}
and extended by Kelly and
Sherman \cite{gks} for the Ising model, which is a special case of
\rf{partition}, \rf{correl} with $n=1$. Ginibre proved the inequalities for the 
XY model, a case with
$n=2$ \cite{ginibre}. Ginibre's method extends to 
other measures for $n=1,2$ 
\cite{glimm, simon}. No one to date has succeeded in proving the second 
G.K.S. inequality for a model of the form \rf{partition}, \rf{correl}
for $n\ge 3$. Significantly, Sylvester showed that
Ginibre's method of proof fails for the $n\ge 3$ classical Heisenberg magnet
(non-linear sigma models)
\cite{sylvester}. {\it We show that both G.K.S. 
inequalities are satisfied for a certain class of models with arbitrary 
$n$}. We are able to find expressions for the measures for some
of these models. As yet, the results have not
been extended to familiar examples, such as 
Euclidean field theories with quartic 
interactions or the classical Heisenberg magnets. Nonetheless, it is significant that it is 
possible to prove the inequalities
for any non-Abelian systems.

A
particular limit of the potential
yields $\mu[s(x)]=\delta[1-s(x)\cdot s(x)]$, \rf{partition} and \rf{correl}, which
is the classical Heisenberg magnet. On a regular hypercubic lattice with
translation-invariant couplings, such models are Euclidean
quantum field theories with $\mu[s(x)]=e^{{-a^{d}\cal V}[s(x)\cdot s(x)]}$, where
$a$ is the lattice spacing for the
potential function $\cal V$ in the Lagrangian. The O($n$) nonlinear sigma model
is the Euclidean quantum field theory which is a classical Heisenberg magnet.

The 
so-called second Griffiths inequality (Theorem 2 below) is proved here
for measures satisfying a particular condition:
\beq
\int d^{n}s\;\mu(s)\;e^{q \cdot s}=e^{\Xi(q\cdot q)}\; \label{transform}
\eeq
for any $q\in {{\rm I}\!{\rm R}}^{n}$ and for 
a function $\Xi(\cdot)$  
which is analytic in $z=q\cdot q$ within a real interval $[0,q_0^{2}]$ for some $q_0\neq 0$, and whose
Taylor expansion in
this interval
has real nonnegative coefficients:
\beq
\Xi(z)=\xi_{0}+\xi_{1}z+\xi_{2}z^{2}+\xi_{3}z^{3}+ \cdots\;\;,\;\;z\in
[0,q_{0}^{2}]\;,\;\; \xi_{1}\ge 0, 
\xi_{2}\ge 0, \xi_{3}\ge 0, \dots \;\;.\label{derivatzero}
\eeq
We note that analyticity is not a particularly strong assumption; it is 
needed for the expansion of correlation functions
(in $J_{xy}$ and $h_{x}$) to converge in a finite volume.

We note that \rf{transform} may be inverted, giving a measure $\mu(\cdot)$ in terms of
any function $\Xi(\cdot)$ satisfying \rf{derivatzero}:
\beq
\int \frac{d^{n}q}{(2\pi)^{n}}\;e^{\Xi(-q\cdot q)}\;e^{-iq \cdot s}=\; 
\mu(s)  \label{inversetransform}
\eeq

If $\Lambda$ is a regular hypercubic lattice of dimension $d$, lattice
spacing $a$ and $J_{xy}=J_{yx}$ is 
\beq
J_{xy}=\left\{ 
\begin{array}{cc}
\frac{1}{2g^2}\;,& {\rm for}\;x,y \;{\rm nearest\;neighbors} \\
r \;,& {\rm for}\;x=y \\
0\;,&{\rm otherwise} 
\end{array}
\right. \;, \label{ferrocoupling}
\eeq
then \rf{partition}, \rf{correl} describes a lattice-regularized 
field theory with bare 
coupling constant
$g$ and external current $h_{x}$. If the measure is of the form
\beq
\mu(s)=e^{-\beta s\cdot s -\gamma_{0} (s\cdot s)^{2}+\cdots}\;,
\label{formofmeasure}
\eeq
the bare mass is
\beq
m_{o}^{2}=(-2d-2rg^{2}+2g^{2}\beta)a^{-2}\;,
\label{baremass}
\eeq 
and the bare quartic interaction is $\lambda_{0}=a^{4-d}g^{4}\gamma_{0}$. This is 
easily seen by
writing 
\beq
\exp{\frac{1}{2g^{2}}\sum_{<x,y>\in \Lambda}s(x)\cdot s(y)}=
\exp -\frac{1}{4g^{2}}\sum_{<x,y>\in \Lambda} \left\{ [s(x)-s(y)]^{2}
-2 \;s(x)^{2} \right\} \;.
\nonumber
\eeq 
\beq
\;\;\;\;\;\;\;\;\;\;\;\;\;\;\;\;\;\;\;\;\;
=\exp  \left\{-\frac{1}{2g^{2}}\;\sum_{x\in \Lambda}\sum_{\nu=1}^{d} [s(x+\nu)-s(x)]^{2}
+\frac{d}{g^{2}} \;\sum_{x\in \Lambda}s(x)^{2} \right\} \;.
\nonumber
\eeq 
There can, of course, be higher-order bare interactions as well.


\section{The G.K.S. inequalities}
\setcounter{equation}{0}
\renewcommand{\theequation}{2.\arabic{equation}}

In this section, we discuss
\newtheorem{axiom1}{Theorem}
\begin{axiom1}
: The correlation functions \rf{correl} of a lattice spin model on a finite
lattice $\Lambda$ with
\end{axiom1}
\beq
\mu(s_{1},\dots, -s_{l},\dots,s_{n} )=
\mu(s_{1},\dots, s_{l},\dots,s_{n} )\ge 0 \;.
\label{symmetry}
\eeq
{\it satisfy the inequality}:
\begin{center}
{\em G.K.S. I}:  $\;\;\;\;\;\;\left< s(x)\cdot s(y)\right> \ge 0\;$, 
\end{center}

\noindent
{\it for all} $x,y\in \Lambda \;\;$.

\vspace{5pt}
\noindent and 

\newtheorem{axiom2}[axiom1]{Theorem}
\begin{axiom2}
: The correlation functions \rf{correl} on a finite lattice $\Lambda$
satisfy the inequality:
\end{axiom2}

\begin{center}
{\em G.K.S. II}: 
$\;\;\;\;\;\;\left<s(x)\cdot s(y)\;s(u)\cdot s(v) \right>-
\left<s(x)\cdot s(y)\right>\left< s(u)\cdot s(v) \right> \ge 0\;$, 
\end{center}

\noindent
{\it for all}
$x,y,u,v\in \Lambda$, {\it provided that the function} $\mu(s)$ {\it satisfies} 
\rf{transform}, \rf{derivatzero}.

\vspace{5pt}

Notice that \rf{transform} is a stronger condition than \rf{symmetry}. The proof
of the first theorem is quite well known and we include it purely for
pedagogical reasons. Though 
both theorems are proved for a finite lattice, they will hold
by continuity in the infinite-lattice limit, provided that the limit
exists.

To prove the first theorem we use the following simple fact:
\newtheorem{axiom3}{Lemma}
\begin{axiom3}
: For any choice of $\mu(\cdot)$ 
satisfying \rf{symmetry}, the 
following integral is nonnegative:
\end{axiom3}
\beq
I^{\alpha_{1},\dots,\alpha_{n}}  =\int d^{n} s\;\mu(s)\;\; 
(s_{1})^{
\alpha_{1}}\cdots(s_{n})^{\alpha_{n}}\ge 0 \;,\nonumber
\eeq
{\it for any nonnegative integers $\alpha_{1},\dots, \alpha_{n}$.}

\noindent
{\it Proof}: The hypothesis is true if any 
$\alpha_{k}$, $k=1,\dots, p$ is
odd; for the integrand changes sign under $s_{k}\rightarrow -s_{k}$, and the 
integral
vanishes by symmetry. Hence we assume that each 
$\alpha_{k}$, $k=1,\dots, n$ is even. Then 
the integral is
\beq
I^{\alpha_{1},\dots,\alpha_{n}} =2^{n}\int d^{n} s\;\mu(s) 
\;\theta(s_{1})\cdots\theta(s_{n})\; (s_{1})^{\alpha_{1}}
\cdots(s_{n})^{\alpha_{n}} \ge 0\;,\label{posint}
\eeq
where $\theta(\cdot)$ is the usual step 
function: $\theta(w)=1$ 
for real $w\ge 0$ and $\theta(w)=0$ for real $w<0$.

With this lemma established it is easy to find the

\vspace{5pt}

\noindent
{\it Proof of Theorem 1}: For a finite lattice $\Lambda$ the 
Taylor expansions
of \rf{partition} and \rf{correl}
in the coefficients $J_{xy}$ and $h_{x}$ is convergent. The 
expansion is a nonnegative linear combination  
of products of integrals of the form \rf{posint}, each of which is
positive or zero. The hypothesis of Theorem 1 
immediately follows.

Theorem 2 is a
deeper result. We need another lemma to prove it.
\newtheorem{axiom4}[axiom3]{Lemma}
\begin{axiom4}
(Ginibre's Inequality): For 
a measure $\mu(\cdot)$, satisfying 
\rf{transform} and \rf{derivatzero}, the
following integral converges and is nonnegative:
\end{axiom4}
\beq
I^{\alpha_{1},\dots,\alpha_{n};\beta_{1},\dots,\beta_{n}}
 =\int d^{n} s\!\! \int d^{n}t\; \mu(s)\mu(t) \;\;\;\;\;\;\;\;\;
\;\;\;\;\;\;\;\;\;\;\;
\;\;\;\;\;
\nonumber
\eeq
\beq
\;\;\;\;\;\;\;\;\;\;\;\;\;\;\;\;\;\;\;\;
\times (s_{1}+t_{1})^{\alpha_{1}}\cdots (s_{n}+t_{n})^{\alpha_{n}}
(s_{1}-t_{1})^{\beta_{1}}\cdots (s_{n}-t_{n})^{\beta_{n}} \ge 0 
 \;,\label{lem2}
\eeq
for any nonnegative integers 
$\alpha_{1},\dots ,\alpha_{n},\beta_{1},\dots ,\beta_{n}$.

\noindent
{\it Proof}: Consider the 
expression
\beq
e^{\Xi[(g+j)^{2}]+\Xi[(g-j)^{2}]} =\int d^{n} s\!\! \int d^{n}t\; \mu(s)\mu(t) 
e^{g\cdot(s+t)}e^{j\cdot(s-t)}
\;. \label{double}
\eeq
The right-hand side of \rf{double} has the form in a neighborhood of $g=0$, $j=0$
\beq
e^{\Xi[(g+j)^{2}]+\Xi[(g-j)^{2}]}
&=& \exp \left(2\xi_{0}+\xi_{1} [(g+j)^{2}+(g-j)^{2}] \right.\nonumber  \\ 
&+& \left. \xi_{2}\{[(g+j)^{2}]^{2}+[(g-j)^{2}]^{2}\}+ \cdots \right) ,\; \label{expr1}
\eeq
where $\xi_{0}=\Xi(0)$. Now 
each term of the Taylor series in the 
exponent is a polynomial in $g^{2}$, $j^{2}$ and $g\cdot j$ with
nonnegative coefficients. For expanding the $l^{th}$ term with the binomial formula yields
\beq
\xi_{l} \sum_{a=0}^{l} \left( \!\!\!\! \begin{array}{cc} & l \\ & a \end{array} \right)
\left[ (-2g\cdot j)^{a}(g^{2}+j^{2})^{l-a}+(2g\cdot j)^{a}(g^{2}+j^{2})^{l-a} \right]\;.
\label{expr2}
\eeq 
All terms with negative coefficients are canceled in this expression. 

Next observe that the right-hand side of \rf{double} is the generating function for\\
$I^{\alpha_{1},\dots,\alpha_{n};\beta_{1},\dots,\beta_{n}}$, as the derivatives of
the right-hand side of \rf{double} with respect to components of $g$ and $j$ 
yield the integrals \rf{lem2}. Hence these integrals
exist by hypothesis. None of the derivatives with respect to 
components of $g$ and $j$ can be negative, by the argument in the previous
paragraph, and the lemma
is proved, if it assumed that $\Xi(z)$ is analytic in a neighborhood of $z=0$.

We should mention at this point that Ginibre's inequality is not true for the
classical Heisenberg measure $\mu(s)=\delta(1-s\cdot s)$, which
does not satisfy the hypothesis of Lemma 2 \cite{sylvester}.

\vspace{5pt}

\noindent
{\it Proof of Theorem 2}: Following Ginibre \cite{ginibre}, we write the 
connected two-point correlation as an expectation value over 
a distribution of two sets of $n$-component spins 
$s(x)$ and $t(x)$, each in $D^{n}$:
\beq
(Z[G])^{2}\left[ \left<s(x)\cdot s(y)\;s(u)\cdot s(v) \right>-
\left<s(x)\cdot s(y)\right>\left< s(u)\cdot s(v) \right> \right]
\nonumber
\eeq
\beq
=\int d[s] d[t] \exp\left\{ \sum_{w,z\in \Lambda} J_{wz} \;[s(w)\cdot s(z)
+\;t(w)\cdot t(z)] + \sum_{w\in \Lambda} h_{w} \cdot [s(w)+t(w)]
\right\} 
\nonumber
\eeq
\beq
\times
[s(x)\cdot s(y)-t(x)\cdot t(y)][s(u)\cdot s(v)-t(u)\cdot t(v)] \;,\label{ginibre}
\eeq
where we have used the
notation
\beq
\int d[s]=\int \left\{\prod_{x\in \Lambda} d^{n}s(x)\; 
\mu[s(x)] \right\}\;. \nonumber
\eeq
We show that the right-hand side of \rf{ginibre} is nonnegative. Note that
\beq
s(x)\cdot s(y)+ t(x)\cdot t(y)
\;\;\;\;\;\;\;\;\;\;\;\;\;\;\;\;\;\;\;\;\;\;\;\;\;\;\;\;
\;\;\;\;\;\;\;\;\;\;\;\;\;\;\;\;\;\;\;\;\;\;\;\;\;\;\;\;\;\;\;\;\;\;  \nonumber
\eeq
\beq
\;\;\;\;\;\;\;\;\;\;\;\;\;\;\;\;\;\;\;\;\;\; =\frac{1}{2}[s(x) + t(x)] \cdot [s(y) + t(y)] 
+
\frac{1}{2}[s(x) - t(x)] \cdot [s(y) - t(y)]\;,
\label{expand1}
\eeq
and
\beq
s(x)\cdot s(y)- t(x)\cdot t(y)
\;\;\;\;\;\;\;\;\;\;\;\;\;\;\;\;\;\;\;\;\;\;\;\;\;\;\;\;
\;\;\;\;\;\;\;\;\;\;\;\;\;\;\;\;\;\;\;\;\;\;\;\;\;\;\;\;\;\;\;\;\;\;  \nonumber
\eeq
\beq
\;\;\;\;\;\;\;\;\;\;\;\;\;\;\;\;\;\;\;\;\;\; = \frac{1}{2}[s(x) - t(x)] \cdot [s(y) + t(y)] 
+
\frac{1}{2}[s(x) + t(x)]\cdot [s(y) -t(y)]\;. 
\label{expand2}
\eeq
Next, we expand the exponential in \rf{ginibre}. Since 
$J_{wz}\ge 0$, $h_{x}\ge 0$, and using
\rf{expand1} and \rf{expand2} we find that the expansion 
is a linear combination
of products of integrals of the form \rf{lem2} with nonnegative 
coefficients. By Lemma 2, the
right-hand side of \rf{ginibre} is nonnegative.

\section{Examples}
\setcounter{equation}{0}
\renewcommand{\theequation}{3.\arabic{equation}}

Measures satisfying \rf{transform} and \rf{derivatzero} can be constructed, at least
in principle, using \rf{inversetransform}. The explicit evaluation
of the Fourier integral is, unfortunately, usually impossible analytically. We note
that the integral \rf{inversetransform} is defined for a polynomial 
$\Xi(z)=\xi_{1}z+\xi_{2}z^{2}+\cdots +\xi_{2l+1}z^{2l+1}$, $\xi_{1}\ge 0$, $\xi_{2}\ge 0$,
$\;\dots,\; \xi_{2l+1}>0$. For the choice $l=0$, the measure is the uninteresting Gaussian
case.

We do not yet know whether systems satisfying \rf{transform}, \rf{derivatzero} display
phase transitions and critical points in the thermodynamic limit. A simple argument
using \rf{inversetransform}
reveals that $\beta\ge 0$ in \rf{formofmeasure}. Consider 
the first and second derivatives of the
measure $\mu(\cdot)$ using \rf{inversetransform} evaluated at $s=0$:
\beq
\left. \frac{\partial}{\partial s_{a}}\mu(s) \right\vert_{s=0}  =-i\int 
\frac{d^{n}q}{(2\pi)^{n}}
\;q_{a}\;e^{\Xi(-q\cdot q)}=0\;,
\nonumber
\eeq
\beq
\left. \frac{\partial^{2}}{\partial s_{a}\partial s_{b}}\mu(s)\right\vert_{s=0}    
=-\frac{\delta_{ab}}{n}  \int \frac{d^{n}q}{(2\pi)^{n}}
\;q\cdot q\;e^{\Xi(-q\cdot q)}  \;.
\nonumber
\eeq
Since $q\cdot q\;e^{\Xi(-q\cdot q)}$ is positive for $q\cdot q>0$, the 
measure has a local maximum at $s=0$. The possibility of $\beta<0$ in
\rf{formofmeasure}
is thereby eliminated. This does not restrict the bare mass to be positive, as there
is the negative contribution in \rf{baremass} proportional to the dimension.


An obvious question is whether the measure of a scalar field theory with only a quartic
interaction can satisfy \rf{transform}, \rf{derivatzero}. We show that for $\beta=0$
\beq
\mu(s)=e^{-\gamma_{0} (s\cdot s)^{2}}\;,
\nonumber
\eeq
the answer is no. Using \rf{transform} to determine $\Xi(\cdot)$ as a power series
in $\gamma_{0}^{-1/2}$, we find from equation \rf{genfunct1} in the appendix
\beq
e^{\Xi(z)}=\frac{\pi^{n/2}\Gamma(\frac{n}{4}) }{2 \Gamma(\frac{n}{2})\; \gamma_{0}^{n/4}}
\left[1+ \frac{ \Gamma(\frac{n}{4}+\frac{1}{2}) }{2n\Gamma(\frac{n}{4}) } \;\gamma_{0}^{-1/2} z
+\frac{1}{32(n+2)}\gamma_{0}^{-1}z^{2}\right.
\nonumber
\eeq
\beq
\left. +\frac{ \Gamma(\frac{n}{4}+\frac{1}{2}) }{192n(n+4)\Gamma(\frac{n}{4}) }  \;
\gamma_{0}^{-3/2}z^{3}
+\cdots
\right]
\nonumber
\eeq
Comparing this series with \rf{derivatzero} gives for the first
few coefficients
\beq
\xi_{1}=\frac{ \Gamma(\frac{n}{4}+\frac{1}{2}) }{2n\Gamma(\frac{n}{4}) }\; \gamma_{0}^{-1/2} \;,
\;\;
\xi_{2}=\left[\frac{1}{32(n+2)}-
\frac{ \Gamma(\frac{n}{4}+\frac{1}{2})^{2} }{8n^{2}\Gamma(\frac{n}{4})^{2} } 
\right] \;\gamma_{0}^{-1}\;,
\nonumber
\eeq
\beq
\xi_{3}=\left[-\frac{ \Gamma(\frac{n}{4}+\frac{1}{2}) }{96n(n+4)\Gamma(\frac{n}{4}) }  
+\frac{ \Gamma(\frac{n}{4}+\frac{1}{2})^{3} }{24n^{3}\Gamma(\frac{n}{4})^{3} }  
\right]\;\gamma_{0}^{-3/2}\;.
\nonumber
\eeq
It is relatively simple to compute these coefficients for even $n$. The author
has checked that 
$\xi_{2}>0$ for $n=2$ (an Abelian case) but  $\xi_{2}<0$ for $n=4,6,8,10,12$. Thus
a simple quartic measure does not satisfy the conditions for our proof
of the second G.K.S. inequality.

An alternative strategy is to start with a suitable choice for $\Xi(\cdot)$ and
determine $\mu(\cdot)$ using \rf{inversetransform}. For the choice $\xi_{1}=\xi_{2}=
\xi_{4}=\xi_{5}=\cdots=0$:
\beq
e^{\Xi(h\cdot h)}=e^{\xi(h\cdot h)^{3}}\;,
\nonumber
\eeq
the formula \rf{genfunct1} in the appendix may be used to show
\beq
\mu(s)= \frac{\pi^{\frac{n}{2}}\Gamma(\frac{n}{6})}{3\Gamma(\frac{n}{2})\xi^{n/6} } 
\left[
1- \frac{\Gamma(\frac{n}{6}+\frac{1}{3}) }{2n \Gamma(\frac{n}{6}) }\; \xi^{-1/3}\;s\cdot s 
+ \frac{\Gamma(\frac{n}{6}+\frac{2}{3}) }{8n(n+2) \Gamma(\frac{n}{6}) } 
\;\xi^{-2/3}\;(s\cdot s)^{2} \right.
\nonumber
\eeq
\beq
\left.
-\frac{1}{288(n+2)(n+4) } \;\xi^{-1}\;(s\cdot s)^{3} +\cdots
\right]\;,
\nonumber
\eeq
which is, as one would reasonably expect, a nonpolynomial interaction. Unfortunately this
measure falls off more slowly than a Gaussian as $\vert s\vert\rightarrow \infty$. For 
field theory, this means that the potential cannot be approximated by a 
Ginzburg-Landau-type expression.

There are other possible choices of $\Xi(\cdot)$. For example, if $n=3$, for 
an integer $c=0,1,2,\dots$
\beq
e^{\Xi_{c}(h\cdot h)}=(-1)^{c}\frac{\partial^{c}}{\partial \omega^{c}}
\frac{1}{(\Omega^{2}-h\cdot h)^{l}} \;,\nonumber
\eeq
implies that
\beq
\mu(s)=\frac{1}{4\pi}\vert s \vert^{c-1} e^{- \Omega \vert s\vert }\;. \nonumber
\eeq
One can check that $\Xi(z)$ has nonnegative derivatives. On a regular lattice with suitably
translation-invariant couplings, such a model is a $3$-component Euclidean
quantum field theory, whose continuum limit has the
Lagrangian (excluding the source term, which depends on $h_{x}$ on the lattice)
\beq
{\cal L}_{c}=\frac{1}{2 g^{2} } (\partial s)^{2}+\frac{m_{0}^{2}}{2 g^{2} }s^{2}+
\omega \vert s\vert+(1-c)\log \vert s\vert \;,
\label{weird}
\eeq
where $\omega$ is the limit of $\Omega a^{-d}$ as $a\rightarrow 0$. These 
3-component field theories have nonpolynomial
interactions (even for $c=1$). Though they are
unconventional, \rf{weird} may be 
interesting models in their own right, as well as for the
reason that they satisfy the 
second G.K.S. inequality. 

\section{Discussion}
\setcounter{equation}{0}
\renewcommand{\theequation}{5.\arabic{equation}}

A major question suggested by the results of 
this paper is whether the method used here can be extended
to models for which $\mu(s)$ falls off faster with large $s\cdot s$ than a Gaussian
distribution. If
not, another approach may be needed to prove the second G.K.S. inequality for such choices
of $\mu(\cdot)$. 

While the conditions \rf{transform}, \rf{derivatzero} are sufficient for the validity
of the second G.K.S. inequality, it is by no means obvious that they are necessary. It 
may be that Ginibre's inequality (Lemma 2.) is satisfied under weaker conditions. 

The classical Heisenberg measure does not satisfy Ginibre's inequality 
as formulated in reference \cite{sylvester}. We would like to suggest, however, that 
the situation
may not be hopeless for this case. To obtain the classical Heisenberg magnet, all that
is needed is that some measure $\mu(\cdot)$ depending on one or more parameters,
satisfies the criteria
\rf{transform}, \rf{derivatzero} and by taking some limit of these parameters, suitably 
adjusting $r$
in \rf{ferrocoupling}, keeping $g$ fixed
\beq
\lim \;e^{(2d+2rg^{2}) s\cdot s}\mu(s)\;\longrightarrow \delta(s\cdot s-1)\;. \nonumber
\eeq
The expansions of correlation functions in the ferromagnetic couplings one
obtains this way are {\em different} from those of Sylvester, who 
takes $\mu(s)=\delta(s\cdot s-1)$ from the outset.

\vspace{10pt}

{\it Acknowledgements}: The author would like to thank Prof. David Tepper and
Mr. Jing Xiao for discussions on
Ginibre's
inequality. 

\section*{Appendix: Expansions for integral transforms}
\setcounter{equation}{0}
\renewcommand{\theequation}{A.\arabic{equation}}

To evaluate explicitly an integral of the form
\beq
Y_{L}(P) =\int d^{n}Q \; e^{Q\cdot P} e^{-\alpha (Q\cdot Q)^{L}}\;,
\label{genfunct}
\eeq
where $P$ and $Q$ are vectors in ${{\rm I}\!{\rm R}}^{n}$ and $L$ is an integer, is
difficult, except as a power series in $P\cdot P$. We shall examine the terms of
this power series in detail in this appendix and see whether it is convergent.

We wish to compute the integrals
\beq
Y_{j_{1},j_{2},\dots,j_{2l}}=\int d^{n}Q \;
Q_{j_{1}}\cdots Q_{j_{2l}}\; e^{-\alpha (Q\cdot Q)^{L}}\;, \label{termsinexp}
\eeq
which occur in the expansion of \rf{genfunct} in the components
of $P$. To evaluate \rf{termsinexp}, we first find the integral
\beq
Z^{l}_{L}=\int d^{n}Q \; (Q\cdot Q)^{l} e^{-\alpha (Q\cdot Q)^{L}}
=
\frac{\pi^{n/2}\Gamma(\frac{n+2l}{2L})}{L\Gamma(\frac{n}{2})}\alpha^{-\frac{n+2l}{2L}}\;,
\label{simpleint1}
\eeq
To obtain this expression, we first integrated over the angles (yielding the standard result
$\int d\Omega=2\pi^{n/2}/\Gamma(n/2)$), then made a change of variable from
$q=\vert Q\vert$ to $\alpha q^{2L}$. Now this integral is obtained by contracting
the indices $j_{1}$ with $j_{2}$, $j_{3}$ with $j_{4}$, $\dots$, $j_{2l-1}$ with $j_{2l}$ with
$Y_{j_{1},j_{2},\dots,j_{2l}}$ by \rf{termsinexp}:
\beq
Z^{l}_{L}=\delta^{j_{1} j_{2}}\cdots
\delta^{j_{2l-1} j_{2l}}Y_{j_{1},j_{2},\dots,j_{2l}}\;,
\label{simpleint2}
\eeq
where we have used the Einstein summation convention. 

By symmetry, $Y_{j_{1},j_{2},\dots,j_{2l}}$ must have the form
\beq
Y_{j_{1},j_{2},\dots,j_{2l}}=A^{l}_{L}\left( \delta_{j_{1}j_{2}}\delta_{j_{3} j_{4}}
\cdots \delta_{j_{2l-1}j_{2l}}+
{\rm all\; other\; contractions\; of\;index\;pairs}
\right),
\label{pairsofint}
\eeq
for some constant $A^{l}_{L}$. There are $(2l-1)!!=(2l-1)(2l-3)\cdots 3\cdot 1$ terms in parentheses on the
right-hand side of \rf{pairsofint}. We shall show that
\beq
\delta^{j_{1} j_{2}}\cdots
\delta^{j_{2l-1} j_{2l}}\left( \delta_{j_{1}j_{2}}\delta_{j_{3} j_{4}}
\cdots \delta_{j_{2l-1}j_{2l}}+
{\rm all\; other\; contractions\; of\;index\;pairs}
\right) \nonumber
\eeq
\beq
\;\;\;\;\;\;\;\;\;\;\;\;\;\;\;\;\;
=\frac{(n+2l-2)!!}{(n-2)!!}=(n+2l-2)(n+2l-4)\cdots (n+2)n\;,
\label{contrform}
\eeq
below. We can thereby evaluate $A^{l}_{L}$ by using \rf{simpleint1}, \rf{simpleint2} and 
\rf{pairsofint}, obtaining 
\beq
Y_{j_{1},j_{2},\dots,j_{2l}}=
\frac{\pi^{\frac{n}{2}}\Gamma(\frac{n+2l}{2L})}{L\Gamma(\frac{n}{2})}
\frac{(n-2)!!}{(n+2l-2)!!}\alpha^{-\frac{n+2l}{2L}}\;\;\;\;\;\;\;\;
\;\;\;\;\;\;\;\;\;\;\;\;\;\;\;\;\;\;\;\;\;\;\;\;\;\;
\nonumber
\eeq
\beq
\;\;\;\;\;\;\;\;\;\;\;\;\;\;\;\;\;
\times \left( \delta_{j_{1}j_{2}}\delta_{j_{3} j_{4}}
\cdots \delta_{j_{2l-1}j_{2l}}+
{\rm all\; other\; contractions\; of\;index\;pairs}
\right)\;.
\label{resultforint}
\eeq

Next we prove \rf{contrform} by induction. The formula is trivial when $l=1$. Let us
suppose that it is established for some particular value of $l$ and 
show it for $l+1$. Consider
\beq
\delta^{j_{1} j_{2}}\cdots
\delta^{j_{2l+1} j_{2l+2}}\left( \delta_{j_{1}j_{2}}\delta_{j_{3} j_{4}}
\cdots \delta_{j_{2l+1}j_{2l+2}}+
{\rm all\; other\; contractions\; of\;index\;pairs}
\right)\;. \nonumber
\eeq
The terms in the parentheses in this expression are of two types: i) Those containing 
$\delta_{j_{2l+1}j_{2l+2}}$ and ii) those which do not. The sum of terms of type i)
are, by hypothesis,  $\frac{(n+2l-2)!!}{(n-2)!!}n$ (the factor $n$ just comes from
the sum over $j_{2l+1}=j_{2l+2}$). The terms of type ii) can be obtained
by considering the left-hand side of \rf{contrform}, {\em uncontracting} a pair
of indices in the set $\{j_{1},\dots,j_{2l}\}$, contracting one member of this
pair with $j_{2l+1}$ and the other member of the pair with $j_{2l+2}$. There are
precisely $2l$ ways to do this, and the sum of terms of type ii) is
$\frac{(n+2l-2)!!}{(n-2)!!}\cdot 2l$. The sum of all terms of type i) and type ii)
is therefore
\beq
\frac{(n+2l-2)!!}{(n-2)!!}(n+2l)=\frac{[n+2(l+1)-2]!!}{(n-2)!!}\;, \nonumber
\eeq
establishing \rf{contrform} for $l+1$.

Putting these results together gives for \rf{genfunct}
\beq
Y_{L}(P) 
=\frac{\pi^{\frac{n}{2}}\Gamma\left(\frac{n}{2L} \right) }
{L\Gamma\left(\frac{n}{2}   \right) \alpha^{\frac{n}{2L}}}+
\frac{\pi^{\frac{n}{2}} (n-2)!!}{L\Gamma(\frac{n}{2})\alpha^{\frac{n}{2L}}
}\sum_{l=1}^{\infty} 
\frac{(2l-1)!! \Gamma(\frac{n+2l}{2L})}{(2l)!(n+2l-2)!!}\alpha^{-\frac{l}{L}}
(P\cdot P)^{l}\;. \label{genfunct1}
\eeq

To test of the validity of \rf{genfunct1} let us take $L=1$. The relations
\beq
r!!=\frac{r!}{2^{\frac{r+1}{2}} (\frac{r+1}{2})!}\;,r\;{\rm odd}\;, 
\nonumber
\eeq
\beq
r!!=2^{\frac{r}{2}}\left( \frac{r}{2} \right) !\;, r\;{\rm even} \;,
\nonumber
\eeq
and
\beq
\frac{\Gamma(\frac{n}{2}+l)}{\Gamma(\frac{n}{2})}=
\frac{(\frac{n}{2}+l-1)\Gamma(\frac{n}{2}+l-1)}{\Gamma(\frac{n}{2})}
=\cdots = \frac{ (\frac{n}{2}+l-1)(\frac{n}{2}+l-2)\cdots (\frac{n}{2}) 
\Gamma(\frac{n}{2})}{\Gamma(\frac{n}{2})  }
\nonumber
\eeq
\beq
\;\;\;\;\;\;\;\;\;\;\;\;\;\;\;\;\;\;=\frac{(n+2l-2)!!}{2^{l}(n-2)!!}\;, \nonumber
\eeq
imply the familiar result:
\beq
Y_{1}(P)
=\pi^{\frac{n}{2}} \alpha^{-\frac{n}{2}} 
\left[ 1+\sum_{l=1}^{\infty} \frac{(2l-1)!!}{2^{l}(2l)!}\alpha^{-l}
(P\cdot P)^{l} \right]
=\frac{\pi^{\frac{n}{2}}}{\alpha^{\frac{n}{2}}}\sum_{l=0}^{\infty} \frac{1}{l!}
\left(\frac{P\cdot P}{4\alpha}\right)^{l}
\nonumber
\eeq
\beq
\;\;\;\;\;\;\;\;\;\;\;\;\;\;\;=\left(
\frac{\pi}{\alpha}\right
)^{\frac{n}{2}}e^{\frac{P\cdot P}{4\alpha}}
\;. \label{genfunct2}
\eeq

To establish convergence of \rf{genfunct1}, we compare the terms in the series for $L>1$ to those for $L=1$. Notice
that for $L>1$,
\beq
\frac{(2l-1)!! \Gamma(\frac{n+2l}{2L})}{(2l)!(n+2l-2)!!}<
\frac{(2l-1)!! \Gamma(\frac{n+2l}{2})}{(2l)!(n+2l-2)!!}\;. \nonumber
\eeq
Since the series for $L=1$ converges, so must that for $L>1$.

\end{document}